# The cylindrical shock at underwater wire explosion

Sergey G. Chefranov and Daniel Maler

Physics Department, Technion-Israel Institute of Technology, Haifa 32000, Israel

csergei@technion.ac.il

daniel.maler@campus.technion.ac.il

*Abstract*

New analytical solution of the piston and shock evolution for the wire electrical explosion in water is obtained. This is provided on the base of the compressible Euler equations without usually a prior introduction of any self-similarity hypothesis. It is shown that diverging cylindrical shock is transformed into acoustic wave in a finite time, even without taking into account of dissipation. The correspondence with experimental data on underwater electrical explosion of thin wire is represented.

## Introduction

The research on shock waves (shocks) generated by either chemical explosions or electrical discharges or by the supersonic propagation of bodies has attracted continuous attention in the fields of hydrodynamics, plasma and space physics, as well as in areas related to thermonuclear research [1-3]. The theoretical studies, conducted in [4-10], considered only self-similar or linear, like in [4, 5], solutions to describe spherical and cylindrical shocks generated by either a sphere or a cylindrical pistons. Moreover, the law of piston evolution in time is also usually a priory artificially determined without using the solution of hydrodynamics equations for it. However, in [11] it is noted that the radius of the cylindrical piston formed as a result of a wire electrical explosion in water increases by the same power law as that obtained in [12] for the self-similar expansion of the cylindrical shock. The latter indicates the possibility of a strong correlation between the piston and shock evolution laws. This correlation can be maintained at significant distances between the piston and the shock, and this correlation may also be responsible for the secondary shocks observed in experiments [11].

Thus, the question of the piston evolution and its influence on the shock parameters was not addressed appropriately when the law of the piston evolution is not obtained from the solution of non-linear hydrodynamic equations without using of any self-similarity hypothesis. It is also important to obtain the analytical solution of this problem because up to now only the numerical solution of hydrodynamic equations [13] is obtained that accounts for the piston's effect on the shock evolution by the energy transferred to the shock during its transition to self-similar propagation.

In this paper, the analytical theory of the evolution of cylindrical piston is presented on the base of generalization of the underwater spherical symmetry explosion theory for incompressible limit [14] to the cylindrical and compressible cases, when the shock evolution is also considered. The corresponding expansion laws for pistons here are not given in advance, as in [5-10], but are determined using the solution of the nonlinear Euler equation for a compressible medium without introduction of any self-similarity hypotheses. In order to determine the shock parameters, the solution obtained for piston evolution is used in the assumption of uniform distribution of the density between the piston and shock. The finite-time decay of the shock to the acoustic wave is also determined and a compare with experimental date is obtained.

In Sect. I the underwater explosion is considered for the cylindrical piston evolution in the compressible case. In Sect. II, the evolution of shock is considered assuming uniform water compression between the piston and shock front. In Sect. III, the comparison of the theory and known experimental date is represented. In Appendix the derivation of compression after shock is provided on the base of compression before the piston.

### I. Cylindrical piston evolution

Let us consider underwater electric explosion of an isolated thin and long wire, when the known isothermal equation of state of water for the compressible case reads [13]:

$$p - p_0 = \frac{K}{n}((\frac{\rho}{\rho_0})^n - 1) \quad (1.1)$$

In (1.1) $p$, $\rho$ are respectively the pressure and density of the water behind the shock front; $p_0$, $\rho_0$ are respectively the pressure and density of the undisturbed water, characterized by the compression module $K = \rho_0 c_0^2 = 2.2 \times 10^9 \, Pa$. Up to the pressure of 25kbar the value of adiabatic index $n = 7.15$ may be useful and $c_0$ is the sound velocity in non-disturbed water at normal conditions.

In addition to Eq. (1.1) the Euler equations for compressible medium in cylindrical variables $(z, r, \varphi)$ are considered in the reference of frame associated with the center of wire, when only radial component of the potential flow velocity is nonzero:

$$\frac{\partial u_r}{\partial t} + u_r \frac{\partial u_r}{\partial r} = -\frac{1}{\rho}\frac{\partial p}{\partial r} \qquad (1.2)$$

$$\frac{\partial \rho}{\partial t} + u_r \frac{\partial \rho}{\partial r} + \rho\left(\frac{\partial u_r}{\partial r} + \frac{u_r}{r}\right) = 0 \qquad (1.3)$$

From (1.2) and (1.1) the equation for the water flow velocity $u_r$ is obtained:

$$\frac{\partial u_r}{\partial t} + \frac{\partial}{\partial r}\left(\frac{u_r^2}{2} + \frac{c_0^2}{(n-1)}\left(\frac{\rho}{\rho_0}\right)^{n-1}\right) = 0 \qquad (1.4)$$

Taking into account that $u_r = (\partial \Phi / \partial r)$, the equation for the potential of the velocity field reads (when an arbitrary function of time, obtaining at integration is considered as zero because potential is also determined only up to an arbitrary function of time)

$$\frac{\partial \Phi}{\partial t} + \frac{1}{2}\left(\frac{\partial \Phi}{\partial r}\right)^2 + \frac{c_0^2}{n-1}\left(\left(\frac{\rho}{\rho_0}\right)^{n-1} - 1\right) = 0 \qquad (1.5)$$

An approach, considered by Milne-Thomson in [14], allows one to obtain the evolution of the spherical gas piston formed as the result of an underwater point explosion and the spatial and temporal dependence for the potential $\Phi$, including that at the piston boundary. But in [14] only the case of an incompressible liquid is considered when in the spherical analog of (1.5) instead of term $\frac{c_0^2}{(n-1)}\left(\frac{\rho}{\rho_0}\right)^{n-1}$ the term $\frac{p}{\rho_{const}}$ is used for water with constant density $\rho_{cons}$. This approach in [14] considers the liquid radial flow induced by an expanding spherical piston, similar to the radial flow formed by a point-like source located at the origin and having time dependence intensity.

When a wire explodes in water, an expanding cylindrical piston boundary is formed and the corresponding shock also has a cylindrical symmetry. Let the cylindrical gas cavity formed by this underwater explosion at time $t = 0$ have radius $R(t=0) = R_0$, length $L$, and pressure $p_{g0}$. At time $t > 0$, the boundary of this cavity radially expands in the surrounding compressible fluid. We consider the evolution of the radius $R(t)$ of the cylindrical cavity, assuming that the length $L$ of the cavity does not change. In the case of adiabatic gas expansion, one obtains

$$\frac{p_1(t)}{p_{g0}} = \left(\frac{R_0^2}{R^2}\right)^g . \qquad (1.6)$$

Here, $p_1(t)$ is the pressure in the cavity at $t > 0$, when $p_1(0) = p_{g0}$.

Let us consider, instead of point-like source of theory [14], a linear-like source located along the axis $z$ which is used for the model of water flow after underwater explosion of the thin wire. The radial velocity of water flow in (1.4), (1.5) is modeled by velocity field potential $\Phi$ which is a fundamental solution of the Helmholtz operator in two-dimensional case [15]:

$$\Delta_2 \Phi - \frac{1}{\lambda^2} \Phi = 2\pi m(t) \hat{\delta}(r) \qquad (1.7)$$

In the right hand side of equation (1.7) is used $\hat{\delta}$ - the Dirac delta-function [15].

The equation (1.7) is valid because for the linear three-dimensional source of intensity $m(t)$ with periodic distribution along axis $z$ the equation for the potential $\Psi$ of velocity field has the form: $\Delta_2 \Psi + \frac{\partial^2 \Psi}{\partial z^2} = 2\pi m(t) \hat{\delta}(r) \exp(iz/\lambda); \Psi = \Phi(r) \exp(iz/\lambda)$ when divergence of velocity is nonzero only on the axis $z$. When the limit of large $\lambda \to \infty$ is considered equation (1.7) is also valid for negligible small values of the velocity field component along axis $z$.

For the case when in (1.7) the characteristic scale along z- axis is the length of the wire $\lambda = L$ the solution of (1.7) is the fundamental solution of two-dimensional Helmholtz operator [15]:

$$\Phi = -m(t)K_0(r/L) \tag{1.8}$$

From boundary condition $(\frac{\partial \Phi}{\partial r})_{r=R(t)} = \frac{dR}{dt} \equiv \dot{R}$ it is possible to obtain from (1.8) the potential in the form

$$\Phi = -\dot{R}LK_0(r/L)/K_1(R/L) \tag{1.9}$$

In (1.9) $K_0; K_1$ are the Macdonald functions of zero and first order respectively.

For simplicity let us consider in (1.8) the limit $R \ll L; r \ll L$ when from (1.8) it is obtained representation (because $K_0(r/L) \approx -\ln(r/2L) + o(1)$ )):

$$\Phi = \dot{R}R\ln(r/2L) \tag{1.10}$$

In (1.10) the boundary condition $(\partial \Phi / \partial r)_{r=R(t)} = \dot{R}$ is also valid as in (1.9).

From (1.1), (1.5) and (1.10) it is possible to obtain equation for radius of cylindrical piston:

$$\frac{c_0^2}{n-1}((1+\frac{n}{\rho_0 c_0^2}(p_w(t)-p_0))^{\frac{n-1}{n}} -1) + \frac{\dot{R}^2}{2} + (\ddot{R}R + \dot{R}^2)\ln(R/2L) = 0. \tag{1.11}$$

Here, we used that $p_1(r=R(t)) = p_w(t)$ determines from (1.6) the water pressure at the cylindrical cavity boundary $r = R(t)$. To solve Eq. (1.11), $p_w(0) = p_{max} = p_{g0}$ should be a known value at $t=0$. Equation (1.11) gives generalization of theory [14] on cylindrical and compressible cases.

Let us introduce a characteristic time $t_0$ as the duration of the explosion, a length $l_0$ as the characteristic scale and the corresponding velocity scale $u_0 = l_0/t_0$. Now, using dimensionless variables $u_0 = l_0/t_0$, $x = R/l_0$, $x_0 = R_0/l_0$, $\tau = t/t_0$ one obtains the dimensionless pressure at the cylinder boundary:

$$p_E = p_{gK}(\frac{x_0}{x(\tau)})^{2g}; p_{gK} = \frac{np_{max}}{K} \quad , \quad (1.12)$$

Now, Eq. (1.11) reads (only for cases with $\alpha x \ll 1$):

$$(x\ddot{x} + \dot{x}^2)\ln(\alpha x) + \frac{\dot{x}^2}{2} + \beta P(x) = 0;$$
$$\alpha = \frac{l_0}{2L}; \beta = \frac{c_0^2}{(n-1)u_0^2}; p_{00} = \frac{p_0 n}{K}; p_E = \frac{pn}{K}. \quad (1.13)$$
$$P(x) = [1 - p_{00} + p_E(x))]^{\frac{n-1}{n}} - 1$$

From (1.9) instead of the equation (1.13) it is possible to obtain more complex equation for the cylindrical piston evolution in the form:

$$x\ddot{x} + \dot{x}^2[1 + \alpha x F_2(\alpha x)] - \beta \alpha x F_1(\alpha x)P(x) = 0;$$
$$F_1 = \frac{K_1(\alpha x)}{K_0(\alpha x)}; F_2 = \frac{1}{F_1} - \frac{F_1}{2} \quad (1.14)$$

In the limit of inertial motion of water (when pressure term in (1.2) is zero, see also [16]) which is realized when $\beta \ll 1$ and the energy, released in the explosion, is $E_{max} \gg E_0 = \rho_0 c_0^5 t_0^3/(n-1)^{5/2}$ if the characteristic length is determined by means of maximal energy $E_{max}$, time of explosion $t_0$ and density $\rho_0$ in the form $l_0 = (E_{max} t_0^2 / \rho_0)^{1/5}$.

The solution of Eq. (1.13) determines the water radial velocity at the cylinder boundary. In the case $u_0^2 \gg c_0^2$, the term with $\beta$ in Eq. (1.13) can be neglected and for the case $|\ln \alpha x| \gg 1; \alpha x \ll 1$ equation (1.13) takes the form:

$$\ddot{x}x + \dot{x}^2 = 0. \quad (1.15)$$

The solution of Eq. (1.15) reads

$$x = x_0\left[1 + \frac{2\dot{x}_0}{x_0}\tau\right]^{1/2}. \quad (1.16)$$

Let us note that in the limit of large times one obtains $R(t) \cong O(t^{1/2})$, which coincides with the self-similar expansion of the cylindrical shock [12].

However, one can argue that at such short times one should take into account the dynamics of the energy density deposition into the exploding wire. The same solution (1.16) is also take place directly from equation (1.14) for any $\beta \leq O(1)$, but in the limit $\alpha \to 0$.

If the value of $\beta$ cannot be neglected, the first integral of Eq. (1.13) gives invariant

$$I = x^2 \dot{x}^2 \ln(\alpha x) + 2\beta \int dx x P(x) = const. \tag{1.17}$$

The representation of invariant for equation (1.14) has more complex form:

$$I_\alpha = \frac{x^2 \dot{x}^2 B(\alpha x)}{2} - \alpha\beta \int dx x^2 F_1(\alpha x) B(\alpha x) P(x); \tag{1.18}$$
$$B(\alpha x) = \exp(2\alpha \int dx F_2(\alpha x))$$

In (1.17) and (1.18) function $P(x)$ is determined from (1.13) and (1.12). From (1.18) in the limit of inertial fluid motion with $\beta \to 0$ it is possible to obtain solution in the form:

$$\int dx x \sqrt{K_0(\alpha x)} \exp\left(\alpha \int dx \frac{K_0(\alpha x)}{K_1(\alpha x)}\right) = \tau \sqrt{I_\alpha(\beta = 0)} + C \tag{1.19}$$

In (1.19) the arbitrary $C = const$ is determined from the initial conditions. Solution (1.19) gives generalization of solution (1.16) on the case of arbitrary finite value of the parameter $\alpha$ which gives the dependence of piston radius evolution from the initial length of exploding wire.

**II. Effect of piston boundary on the shock evolution**

1. Let us consider the modification of equations (1.6) and (1.12) by taking into account the time dependence of the power $W(t)$ deposited into the cavity in the equation of energy balance:

$$\eta W(t) = \frac{d}{dt}(\varepsilon V) + p \frac{dV}{dt}. \tag{2.1}$$

Here, $\eta < 1$ is the coefficient that determines the efficiency of the energy deposition and $\varepsilon$ is the internal energy density absorbed in the volume $V$ of the cavity formed by the explosion. Now, using also the equation of state for ideal gas $p = (g-1)\varepsilon$, that instead of equation (1.13) it is possible to obtain system of equations that gives the generalization to (1.13) which is taken into account equation (2.1) with new representation for function $P(x) \equiv \delta^{n-1}$ in invariant (1.17):

$$\dot{x} = z;$$
$$\dot{z} = -\frac{1}{x \ln \alpha x}(\beta \delta^{n-1} + z^2(\frac{1}{2} + \ln \alpha x));$$
$$\delta = \left[1 - p_{00} + \frac{(p_{00}x^{2g}(0) + y(\tau))}{x^{2g}}\right]^{1/n}, 0 \leq \tau \leq 2; y = px^{2g}n/K_0$$
$$\delta = \left[1 - p_{00} + \frac{1}{x^{2g}}\left(p_{00}x^{2g}(0) + y(2)\right)\right]^{1/n}; \tau \geq 2; \quad (2.2)$$
$$\dot{y} = x^{2(g-1)}f(\tau); y(0) = 0$$
$$f(\tau) = d\exp(-\alpha_1(1.407 - \tau)^2); d = \frac{n\eta(g-1)t_0 W_{max}}{\pi L l_0^2 K}; \alpha_1 = 5.5677, 0 < \tau < 2;$$
$$f(0) = f(2) = 0$$

The form of function $f(\tau)$ is near the same as in experiment [17], where for electrical explosion of wire (Cu or Al) $E_{max} = t_0 W_{max} \approx 2900 J; t_0 \approx 10^{-6} \sec; L \approx 4 \times 10^{-2} m$. For estimations it is also possible to use in (2.2) value $d \approx 200$, which is corresponding to values $\eta \approx 0.8; g = 4/3; l_0 \approx 10^{-3} m; p_{00} = 3.25 \times 10^{-4}; \beta \approx 0.06; K \approx 2.2 \times 10^9 Pa$.

The results of solution to the system (2.2) are represented in Fig.1 and Fig.2.

Equation (2.2) may be considered instead of the corresponding generalization of more complex equation (1.14) only in the case when the limit $\alpha x \ll 1$ is valid. For the estimation of the value of parameter $\alpha$ it is possible to consider the case when $l_0 \approx 10^{-3} m; L \approx 4 \times 10^{-2} m$ that gives estimation $\alpha \approx 1/80$. So the condition $\alpha x \ll 1$ is valid up to values $x \leq 10$.

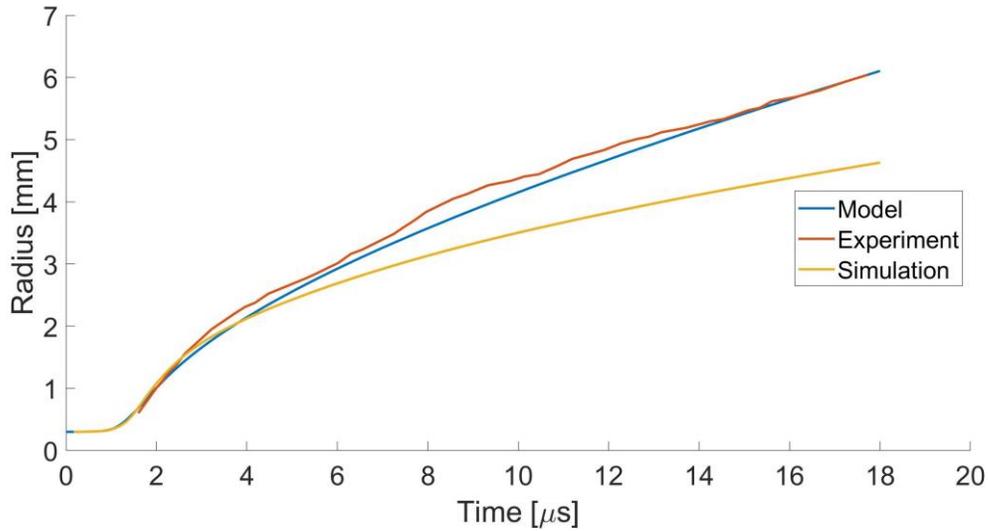

Fig. 1

Radial expansion of wire according to model (2.2) at $\alpha_1 = 5.5677; d = 200$, as compared with simulation and experiment [17].

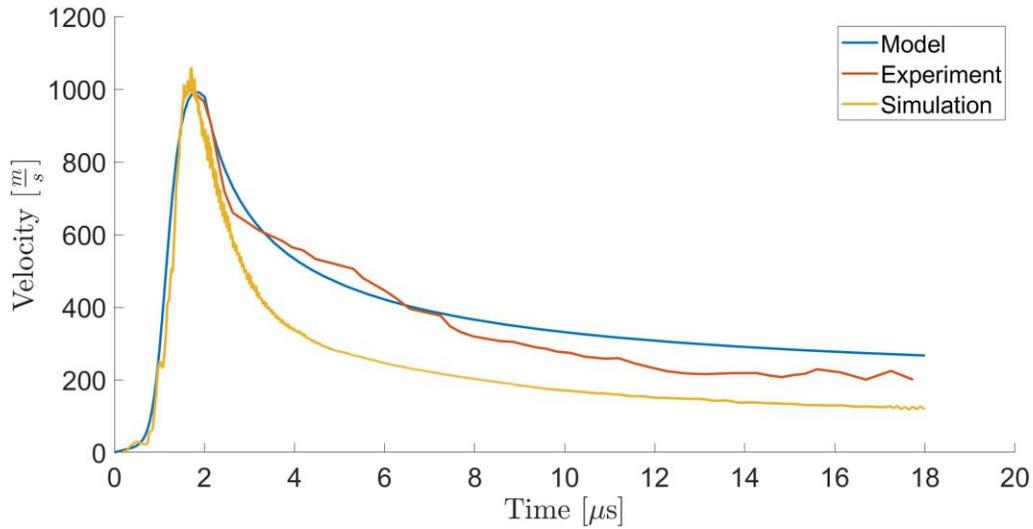

Fig.2 The velocity of piston (expanding wire) according to the model (2.2), as compared with simulation experiment [17]

2. On the base of (2.2) it is possible to obtain the evolution of the dimensionless shock radius $x_D = R_D / l_0$ and shock velocity $D = dR_D / dt$:

$$D/c_0 = \beta_1 \frac{dx_D}{d\tau} = \left[\frac{\delta_D(x_D,x)(\delta_D^n(x_D,x)-1)}{n(\delta_D(x_D,x)-1)}\right]^{1/2}; \beta_1 = \frac{l_0}{t_0 c_0} = 2/3;$$

$$\delta_D(x_D,x) = \delta(x) - \frac{\dot{\delta}}{z}(x_D - x) + O((x_D - x)^2); \quad (2.3)$$

$$x_D(0) = x(0)$$

In (2.3) it is used the representation of compression after shock in the form of function $\delta_D(x_D,x)$, where the compression before piston $\delta(x)$ (see Appendix).

In the limit cases it is also possible to assume the suggestion of the radial homogeneity of compression in the region between piston and shock, when $\delta_D = \delta$ at the condition:

$$x_D - x << \frac{z\delta}{\dot{\delta}} = \left(\frac{1}{\delta}\frac{d\delta}{dx}\right)^{-1} \quad (2.4)$$

From the other side, when (2.4) is not valid it is possible use the representation of $\delta_D$ in (2.3) or of more common representation for it in the form (Appendix):

$$\delta_D = \delta\left(1 + a_1(x_D - x) - a_2(x_D - x)^2 + O((x_D - x)^3)\right);$$

$$a_1 = -\frac{\dot{\delta}}{z\delta}; a_2 = \frac{1}{2z\delta}\left(\dot{\delta}\left(\frac{2}{x} + \frac{\dot{z}}{z^2}\right) - \frac{\ddot{\delta}}{z}\right) \quad (2.5)$$

It is need to use (2.5) in (2.3) instead of representation of $\delta_D$ in (2.3) if the next condition is not valid:

$$x_D - x << a_1 / |a_2| \quad (2.6)$$

Thus the representation for $\delta_D$ in the form which is used in (2.3) is valid only at condition (2.6).

In the case when (2.6) is not valid it is need to use (2.5) in (2.3) or it is possible to use the nonlinear transformation of (2.5) in the known form [18-20]:

$$\delta_D = \delta\left[\frac{a_1 + (a_1^2 + a_2)(x_D - x)}{a_1 + a_2(x_D - x)}\right] \quad (2.7)$$

In the case when $a_2 > 0$ from (2.7) in the limit $x_D - x \gg 1$ it is possible to obtain estimation:

$$\delta_{D\lim} \cong \delta(1 + \frac{a_1^2}{a_2}) \qquad (2.8)$$

3. For simplicity let us suppose the case with $\delta_D \approx \delta$ in (when (2.4) and (2.6) are valid as necessary conditions) and also consider the asymptotic solution (1.16) when is valid in (2.2) for $\tau \geq 2$:

$$\delta = (1 + \varepsilon(\tau))^{1/n}; \varepsilon = \frac{p_{K\max}}{x^{2g}} - p_{00};$$
$$p_{K\max} = y(2) + p_{00} x^{2g}(0) \qquad (2.9)$$

Eq. (2.9) dictates that the shock exists only during a time interval when $\varepsilon > 0$. So at the time $\tau \to \tau_{\max}$ when $x \to x_{\max}$ in (2.9) $\varepsilon \to 0$ where:

$$x_{\max} = \left(\frac{p_{K\max}}{p_{00}}\right)^{1/2g} \qquad (2.10)$$

At $\tau = \tau_{\max}$ and $x = x_{\max}$ the shock turns into an acoustic wave (for $\varepsilon = 0$ the compression becomes $\delta = 1$). For example, in (2.10) $x_{\max} \approx 45$ when $p_{\max}(\tau = 2) = 2.5 \times 10^9 \, Pa$ in (2.10) $p_{K\max} = \frac{p_{\max} n}{K} \approx 8.125; p_{00} = \frac{p_0 n}{K} \approx 3.5 \times 10^{-4}$ for $n = 7.15; g = 4/3$.

**Appendix: Radial distribution of compression**

Now let us consider evolution of compression $\delta = \rho / \rho_0$ on the base of continuity equation (1.3) with radial velocity giving by potential (1.10) in the dimensionless form when $x_1 = r/l_0 \neq 0$ and the last term in (1.3) is zero (because for delta-function is valid [15] $\hat{\delta}(r) = 0, r \neq 0$):

$$\frac{\partial \delta}{\partial \tau} + \frac{x\dot{x}}{x_1} \frac{\partial \delta}{\partial x_1} = 0 \qquad (A.1)$$

Let us consider (A.1) for the determination of terms in the Taylor series:

$$\delta(x_1,\tau) = \delta(x(\tau)) + \frac{(x_1-x)}{1!}\left(\frac{\partial \delta}{\partial x_1}\right)_{x_1=x} + \frac{(x_1-x)^2}{2!}\left(\frac{\partial^2 \delta}{\partial x_1^2}\right)_{x_1=x} + O((x_1-x)^3) \qquad (A.2)$$

This gives the possibility to obtain of the value of compression directly after the shock $\delta_D = \delta(x_1 = x_D)$.

Thus, from (A.1) it is possible to obtain:

$$\left(\frac{\partial \delta}{\partial x_1}\right)_{x_1=x} = -\frac{\dot\delta}{z}; z = \dot x; \dot\delta = \partial \delta / \partial \tau \qquad (A.3)$$

$$\left(\frac{\partial^2 \delta}{\partial x_1^2}\right)_{x_1=x} = -\frac{1}{z}\left[\dot\delta(\frac{2}{x}+\frac{\dot z}{z^2}) - \frac{\ddot\delta}{z}\right] \qquad (A.4)$$

From (A.3) and (A.4) it is possible to obtain (2.5).

Also from (A.3) the boundary condition for the radial derivative of the compression parameter $\delta$ reads when (1.12) is taken into account:

$$(\frac{\partial \delta}{\partial r})_{r=R(t)} = -\frac{1}{l_0}\frac{d\delta(x)}{dx} = \frac{2g}{l_0}\frac{p_{g0}}{Kx\delta^{n-1}}(\frac{x_0}{x})^{2g} \qquad (A.5)$$

**Conclusions**

The new theory is considered for explosion shocks without applying the self-similar shock propagation approach when the piston evolution is determined directly from the exact solution of the Euler equations. Nonlinear hydrodynamic equations of a compressible medium were used to account for the effect of the piston boundary on the shock parameters. Assuming uniformity of the water compression inside the layer between the piston and the shock front, the solutions of the hydrodynamic equations allows one to obtain the relation between the piston and the shock propagations with finite time of shock existence. In future, it is interesting to consider the possibility of refusing this assumption on the base of exact not only potential, but also vortex solutions of three dimensional hydrodynamic equations, which are obtained in [16].